\documentclass[journal, 10pt]{IEEEtran}
\ifCLASSINFOpdf
\else
\fi
\usepackage{cite}
\usepackage{url}
\usepackage{hyperref}
\usepackage{graphicx}
\usepackage[small]{caption}
\usepackage{subcaption}
\usepackage{color,soul}
\usepackage{array}
\usepackage{amsmath}
\usepackage{mathtools}
\usepackage{booktabs}
\usepackage{footnote}
\makesavenoteenv{tabular}
\makesavenoteenv{table}
\usepackage{color}
\usepackage{amssymb}
\usepackage[linesnumbered,ruled,vlined]{algorithm2e}
\usepackage[noend]{algpseudocode}
\usepackage{setspace}

\begin{document}
%
\title{A Graph Partitioning Algorithm with Application in Synthesizing Single Flux Quantum Logic Circuits}

\author{  Ghasem Pasandi and Massoud Pedram\\
  Department of Electrical Engineering-Systems\\ University of Southern California, Los Angeles, CA 90089.

\thanks{
}

}

%

\maketitle

\begin{abstract}
In this paper, a new graph partitioning problem is introduced. The depth of each part is constrained, i.e., the node count in the longest path of the corresponding sub-graph is no more than a predetermined positive integer value $p$. An additional constraint is enforced such that each part contains only nodes selected from consecutive levels in the graph. The problem is therefore transformed into a Depth-bounded Levelized Graph Partitioning (DLGP) problem, which is solved optimally using a dynamic programming algorithm. As an example application, we have shown that DLGP can effectively generate timing-correct circuit solutions for Single Flux Quantum (SFQ) logic, which is a magnetic-pulse-based, gate-level pipelined superconductive computing fabric. Experimental results confirm that DLGP generates circuits with considerably lower path balancing overheads compared with a baseline full-path-balancing approach. For example, the balancing overhead (a critical measure of quality metric) for the SFQ circuit realization in terms of D-Flip-Flop count is reduced by  $3.61 \times$ on average for 10 benchmark circuit, given $p$=$5$.
\end{abstract}
\IEEEpeerreviewmaketitle

\section{Introduction}
\label{Intro:sec}
\setstretch{0.93}
Graph partitioning (GP) consists of dividing nodes of a graph into smaller (typically, equal size) parts to minimize a cost function subject to some constraints. GP plays an important role in many different applications including parallel processing, processing complex networks, image processing, and VLSI design \cite{bulucc2016recent}. Partitioning large VLSI circuits has an important impact on placement, routing, and testability of those circuits. A good partitioning can result in lower total wire-length, which has direct impact on reducing the critical path delay of the circuit and total area of the chip. Moreover, in hardware simulation and test, a good partitioning solution can reduce the number of required multiplexers for passing inter-block signals to the bus architecture of the hardware simulator \cite{alpert1995recent}. For GP problems in VLSI, a circuit graph is generated by considering gates or modules as nodes of the graph and wires connecting them as edges of the said graph.

In this paper, we present the Depth-bounded Levelized Graph Partitioning (DLGP) problem as a new GP problem and provide optimal solution for it using a Dynamic Programming (DP) algorithm. Thanks to DLGP, an important problem in the design flow of the superconductive, Single Flux Quantum (SFQ) circuits is addressed. SFQ gates with switching delay of $1ps$ and energy consumption of $10^{-19}J$ per switching are considered as potential candidates for achieving super high-performance and ultra energy-efficient systems \cite{chen1999rapid,holmes2013energy,likharev1991rsfq}. Despite the superiority of SFQ gates in achieving super fast and ultra energy-efficient circuit realization, there are  a few challenges in their design flow. 

In a SFQ design most circuit elements (including all logic cells) are clocked elements, i.e., each logic cell becomes a pipeline stage. For correct operation of a SFQ logic cell, it is necessary for different inputs of the cell to arrive at the cell input at the same clock cycle. Hence, when different inputs take different path lengths to traverse they must be explicitly path balanced using D-flip-flops (DFFs). 
In this paper, we will address path balancing overhead as an important challenge in the SFQ circuit design process. 
\section{Preliminaries}
\label{sec:pre_prior}
\subsection{Background on Graph Partitioning}
\label{subsec:GPP}
\textit{Definition}: Given a graph $G=(V,E)$, with non-negative edge weights, $w : E \rightarrow \Bbb R^+$, and a size $s(v)$ for each vertex $v \in V$, the \textit{graph partitioning problem (GPP)} is defined as dividing set $V$ into subsets $V_1, V_2, ..., V_K$ such that Eqs. \ref{union_eq1}, \ref{union_eq2} hold and an objective function (read below) is minimized.
\begin{equation}
\label{union_eq1}
V_1\cup V_2 \cup ... \cup V_K = V
\end{equation}
\begin{equation}
\label{union_eq2}
V_i \cap V_j = \emptyset ~~\forall i \neq j
\end{equation}
The above problem is called \textit{K-way} partitioning as well. Size of part $i$ is denoted by $|V_i|$, i.e. $\sum_{v \in V_i} s(v) = |V_i|$. The bounded-size GPP is defined as a GPP problem in which size of the $i^{th}$ part is bounded by $B_i$ ($|V_i| ~\leq B_i$). A special case is \textit{balanced partitioning}, where the size of all parts should be equal modulo a correction factor. More precisely, Eq. \ref{balanced_part} should hold:
\begin{equation}
\label{balanced_part}
\forall i \in \{1, 2, ... , K\}, ~|V_i| \leq \frac{1+\epsilon}{K} \times |V|
\end{equation}
This problem is commonly denoted as a ($K$, $1+\epsilon$)-balanced partitioning problem. If $\epsilon = 0$, the problem is called perfect partitioning, and the special case of $K$=$2$ and $\epsilon$=$0$ is called the \textit{minimum bisectioning}.

As mentioned before, in GPP an objective function should be minimized. The most used objective function for GPP is the \textit{total cut size}, which is defined by the following equation:
\begin{equation}
\label{cost_func1}
\sum_{e \in C} w(e)
\end{equation}
\begin{equation*}
C = \{e=<u,v> \in E : u \in V_i, v \in V_j, i \neq j\}
\end{equation*}
Level of a node $n$ in a graph $G$ is defined as the length of the longest path in terms of the node count from primary inputs of $G$ to node $n$; if  nodes of the graph are logical gates, it is called the logic level. Depth of a graph is defined as the highest level among all nodes in the graph. Depth of a part in a partitioning problem is the difference between the highest and the lowest levels among nodes of this part plus 1. 
 
\subsection{Background on SFQ}
\label{BG-RSFQ:sec}
SFQ gates are pulsed-based and the presence and absence of a pulse are considered as ``1" and ``0", respectively. A pulse is a single quanta of magnetic flux ($\Phi_0$=$h/2e$=$2.07mV\times ps$) with a duration of a few $ps$ and amplitude of a few $mV$. 
In the following, some key properties of SFQ circuits are explained.
\begin{figure}[t]
\centering
\includegraphics[width=0.44\textwidth]{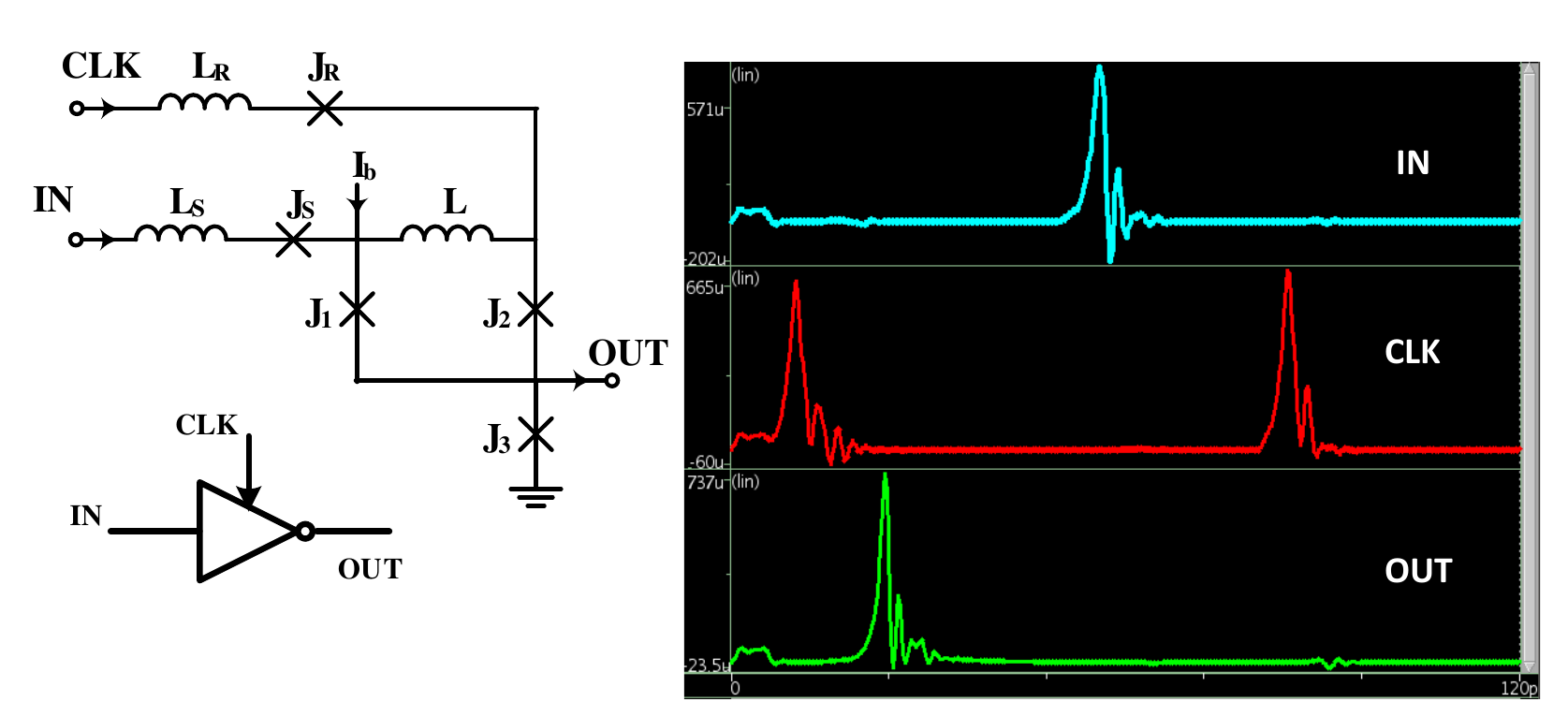}
\caption{Schematic of an SFQ inverter and its waveforms.}
\label{RSFQ_inv}
\end{figure}
\subsubsection{Gate-level pipeline}\label{pipeSubSubSec} SFQ gates (except for confluence buffers, splitters, I/O cells, and T-Flip-Flops) need to receive a clock signal and their operation is synchronized by the clock.
Fig. \ref{RSFQ_inv} shows the circuit diagram of an SFQ inverter and the corresponding waveform to show its functionality. As seen, after the clock pulse comes, when there is no input pulse (which means ``0"), a pulse is generated at the output of the gate representing a ``1". On the other hand, when there is an input pulse, no pulses are generated at the output, meaning a ``0".

\subsubsection{Path Balancing}\label{PathBalancingSubSubSec} 
\begin{figure}[t]
        \centering
         \begin{subfigure}[!t]{0.29\textwidth}
                \centering
                \includegraphics[width=\textwidth]{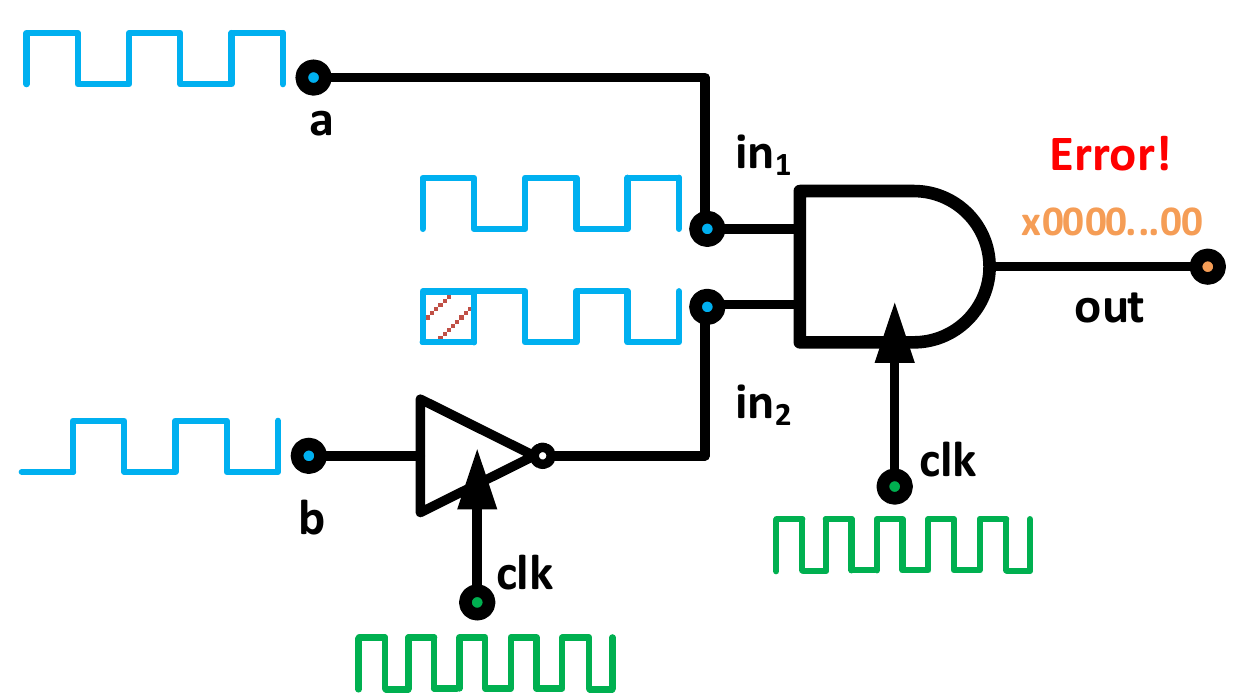}
                \caption{}
                \label{not_balanced_fig}
        \end{subfigure}
        \centering
         \begin{subfigure}[!t]{0.29\textwidth}
                \centering
                \includegraphics[width=\textwidth]{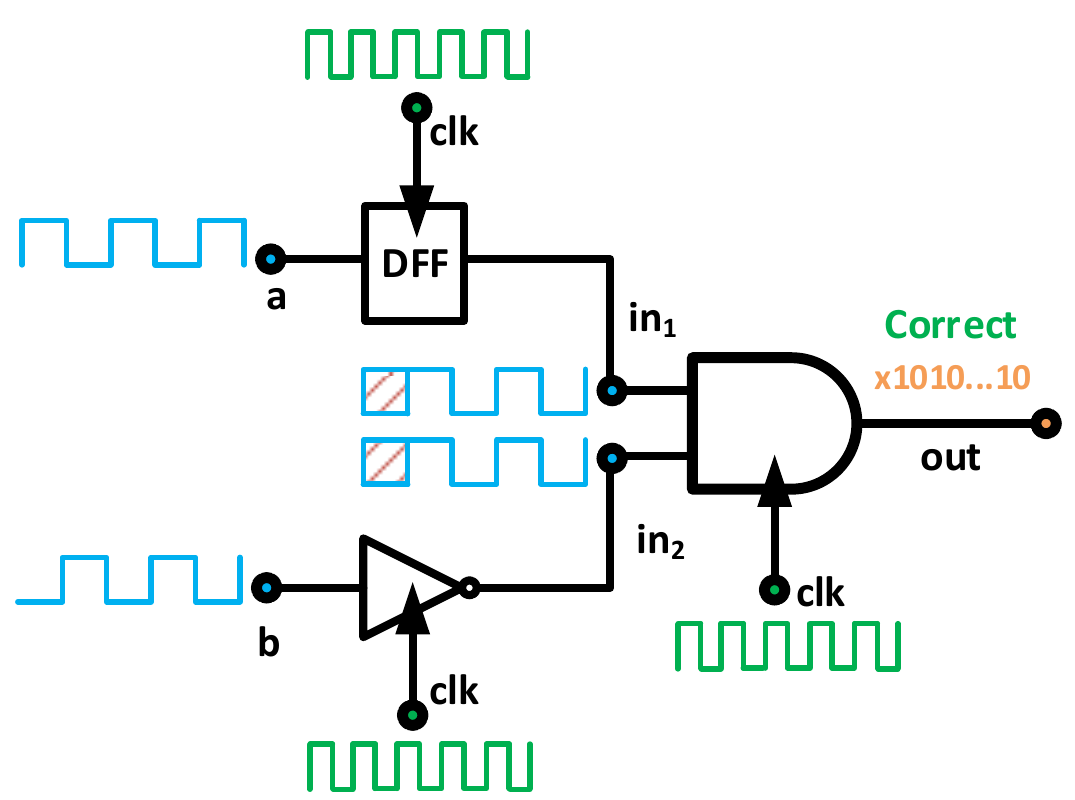}
                \caption{}
                \label{balanced_fig}
        \end{subfigure}
        \caption{Gate-level schematic of Example 1 for showing the necessity of path balancing for correct operation in SFQ circuits.}
        \label{balancing_AND}
\end{figure}

In standard SFQ circuit design, to guarantee the correct operation, all fanins of a gate should have the same logic level. Otherwise, some path balancing DFFs should be inserted into shorter paths. This is called path balancing. Here is an example of necessity of path balancing in SFQ circuits.\\
\textit{Example 1}: suppose that there is a digital signal $a_i$=1010...10 and we want to AND it with invert of another digital signal $b_i$=0101...01. The correct output is: x1010...10. The first bit in the output is not valid because in the first clock, second input of the AND gate ($in_2$) is unknown. Without path balancing, generated values at the output of the AND gate will be x0000...00 which is not correct (Fig. \ref{not_balanced_fig}). The error occurred because the signals on $in_2$ are one level behind the signal on $in_1$. By inserting a path balancing DFF, all fanins of the AND gate will have the same logic level. In the path balanced circuit, as shown in Fig. \ref{balanced_fig}, the correct sequence of bits are generated at the output of the AND gate.  

\subsubsection{D-Flip-Flops (DFFs) in SFQ}\label{FFs_SFQ} There are two types of DFFs in SFQ: Destructive Read Out (DRO), and Non Destructive Read Out (NDRO). In DROs, after reading the internal data of the DFF, it will be destroyed and cannot be read until another value is written into it. In NDRO, read operation will not destroy the stored data in the DFF.  
\section{Prior Work}
\label{sec:prior}
The balanced min-cut k-way partitioning problem is NP-complete \cite{garey1974some, hyafil1973graph}. It remains NP-complete even for $K$=$2$ and with identical vertex size and unit edge weight \cite{garey1974some, lewis1983computers}. 
The first well-known heuristic for solving 2-way balanced partitioning problem is Kernighan-Lin heuristic (K-L) \cite{kernighan1970efficient}. In K-L, a pairwise interchange process is performed by exchanging vertex pairs that yield the largest decrease in the cut-size. The exchanged vertices are locked and the process continues until all vertices are fixed. A well-known modification of K-L heuristic is presented by Fiduccia and Mattheyses in \cite{fiduccia1988linear}. Fiduccia and Mattheyses heuristic (F-M) handles unbalanced parts, supports hyper-edges, and has a linear run-time in total number of circuit pins. After F-M, many other papers presented effective partitioning heuristics using simulated annealing \cite{johnson1989optimization}, multilevel approach \cite{alpert1998multilevel}, network flow \cite{liu1998circuit,liu1998network}, spectral methods \cite{hagen1992new, mcsherry2001spectral} or a unified approach by combining a few methods \cite{liu1998network, lafon2006diffusion}. 

In \cite{lawler1969module}, a module labeling and clustering algorithm is presented which is known as \textit{Lawler's clustering algorithm}. In Lawler's clustering algorithm, an optimal solution for tree clustering to minimize delay with constraints on cluster size and the maximum pin number of each cluster is given. A unity delay model is used for each cluster and it is assumed that delays of interconnects are zero. Lawler's algorithm provides optimal solution for DAGs if replication of modules are allowed. Rajmohan \textit{et al.} presented a clustering algorithm for minimizing the delay in combinational circuits \cite{rajaraman1995optimum}. Unlike Lawler's algorithm, in \cite{rajaraman1995optimum} a general delay model is used and an optimal solution subject to area capacity constraints is given. In \cite{bulucc2016recent,alpert1995recent,hauck1997evaluation, chen2000tutorial}, good surveys on GP and its applications are given.

Different from other GPPs, in Depth-bounded Levelized Graph Partitioning (DLGP) problem, there is a constraint on the depth of each part while balancing each part in terms of total size is not of an interest. In the standard definition, DLGP is considered as un-balanced partitioning. However, since in DLGP the depth of parts should be the same, it can be called as depth-balanced or depth-limited partitioning. In this context, the standard balanced partitioning problem can be called size-balanced partitioning problem.
\section{Problem Formulation and Proposed Algorithm}
\label{sec:prop_alg}
\textit{Depth-bounded Levelized Graph Partitioning (DLGP) problem definition:} Given a directed acyclic graph $G=(V,E)$, a mapping function $\Lambda$ which specifies the level of each node in $V$ to be between 1 and a maximum value of $L$ (i.e., the longest node distance from any source node of $G$), and a positive integer $p$, partition set $V$ into $K$ parts $V_1,V_2,...,V_K$, each of which giving rise to an induced sub-graph $G_1,G_2,...,G_K$, such that (i) the depth of each sub-graph is no more than $p$,  and (ii) if a node of level $l$ is included in some part $V_k$, then all nodes of level $l$ also belong to part $V_k$ . Furthermore, the total cut\_size (TCS) as defined below is minimized:
\begin{equation}
TCS = \sum_{n=1}^{L-1} cut\_size(<Pre(n), Post(n)>)
\end{equation}
where $Pre(n)$ denotes all nodes $i$ in $G$ belonging to parts $V_n'$ such that $n' < n$, $Post(n)$ denotes all nodes $j$ in $G$ belonging to parts $V_n''$ such that $n'' \geq n$. $<Pre(n), Post(n)>$ denotes the edge separator between levels $n$-$1$ and $n$ in $G$ i.e., the set of edges in $G$ which originate at any node $i$ in $Pre(n)$ and terminate at any node $j$ in $Post(n)$. $cut\_size(<Pre(n), Post(n)>$ calculates the number of edges in $G$ that exist between $Pre(n)$ and $Post(n)$ sets.

Next, we define a weighted directed chain graph $C=(U,F,w)$ with nodes labeled $1...L$ where $L$ is equal to the depth of graph $G=(V,E)$. Each such node represents all nodes of $G$ which are at the same level. There will be a directed edge $uv$ in $F$ between nodes $u$ and $v$ only when $v$=$u$+$1$. Weight of the incoming edge to node $v$ accounts for total number of edges connected to any node with level $\geq v$ from nodes with level $<v$. More precisely, if $v$=$u$+$1$, the weight $w_{uv}$ of the $uv$ edge is defined as the number of directed edges in $G$ that connect any nodes in $\{Pre(u),u\}$ and any other nodes in $Post(v)$. If $v\neq u$+$1$, $w_{uv}$=$0$.

Note that the above definition and weight assignment function work equally well for hyper-graphs and hyper-edges (simply add ``hyper" before any occurrence of ``graph" or ``edge".) A directed hyper-edge is one with a distinguished connected node called a source node, thereby, establishing a clear sense of directionality between the source node and all other sink nodes connected by the hyper-edge. 

\textit{Depth-bounded Chain Graph Partitioning (DCGP) problem definition}: Given a weighted chain graph $C=(U,F,w)$ and a positive integer $p$, partition set $U$ into $K$ parts $U_1,U_2,...,U_K$ such that (i) the depth of each part is no more than $p$, (ii) the total cut weight as defined below is minimized: 
\begin{equation}
\label{TCW_eq}
TCW = \sum_{k=1}^{K-1} cut\_weight(<Pre(k), Post(k)>)
\end{equation}
where $Pre(k)$ denotes all nodes $i$ belonging to parts $U_k'$ in $C$ such that $k' < k$, $Post(k)$ denotes all nodes $j$ belonging to parts $U_k''$ in $C$ such that $k'' \geq k$. $<Pre(k), Post(k)>$ denotes the edge separator of $Pre(k)$ and $Post(k)$ in $C$ i.e., the set of edges in $C$ which originate at any node $k'$ in $Pre(k)$ and terminate at any node $k''$ in $Post(k)$. $cut\_weight(<Pre(k), Post(k)>)$ is calculated as the sum of the edge weights $w_{uv}$ of every distinct pair of nodes, $u \in Pre(k)$ and $v \in Post(k)$.

\textbf{Lemma 1}: DCGP and DLGP are equivalent problems, i.e. solving the DCGP problem yields the solution of the DLGP problem and vice versa.

\textbf{Proof}: It is enough to show that a solution to DCGP problem can be transformed into a valid solution for DLGP problem and vice versa. The proof is straight-forward and follows easily from the way the chain graph and $Pre$ and $Post$ sets are defined. It is omitted here to save space. $\blacksquare$ 

Note that when $p$ is equal to or larger than the depth of graph $G$, then there will be only one part equal to $G$ itself and the problem is trivial. 

\textbf{Lemma 2:} The DCGP problem can be solved optimally using the DP algorithm. 

To use DP for finding the optimal solution for the DCGP problem, we define $O(i)$ as the partitioning solution for a sub-graph $G_i$ of the original graph $G$ which minimizes the total cut weight as defined by Eq. \ref{TCW_eq}. Notice that $G_i$ is an induced graph obtained from $G$ by including all nodes of $G$ with levels less than or equal to $i$. $O(L)$ denotes the optimal solution for our problem. 
We initialize $OPT(i)$=$0$ for all $1 \leq i \leq p$. Next the value of the optimal solution, $OPT(i)$, which is defined as the minimum value for $TCW$ for induced graph $G_i$, is calculated recursively as follows:
\begin{equation}
\label{opt_i_eq}
OPT(i) = min_{q} \{ OPT(i-q)+ 
\end{equation}
\begin{equation*}
cut\_weight(<Pre(i-q+1),Post(i-q+1)>) \} 
 ~~for ~~1\leq q \leq p 
\end{equation*}
{\small
}
\textbf{Proof of optimality}: It should be shown that the optimal solution of a subset of problem $O(i)$ is built of optimal solutions for its sub-problems. For this purpose, we use the induction hypothesis as follows: suppose that the $i^{th}$ instance of the problem with optimal solution $O(i)$ has a sub-problem $i-q$ with optimal solution $O(i-q)$ and optimal value of $OPT(i-q)$ = $M$. Suppose that $O(i)$ is built of a solution for  $(i-q)^{th}$ sub-problem with value $M'> M$. Let's call this solution $O'(i-q)$. Now, we can generate another solution for the $i^{th}$ instance of the problem by replacing $O'(i-q)$ with $O(i-q)$. Since $M < M'$, then the new solution for the $i^{th}$ instance of the problem is better than the first one which is a contradiction, because the first solution was supposed to be the optimal solution. Therefore, the optimal solution for $i^{th}$ instance of the problem is built of the optimal solutions for its sub-problems $\blacksquare$.

\textbf{Theorem}: The DLGP problem can be solved optimally.\\
\textbf{Proof}: Using lemma 1 and lemma 2, the proof is straight-forward.

After finding the optimal solution, parts can be generated by tracing the $O(L)$ solution back as follows: Generate an empty set of selected levels $N_{sel}$. Add the indices of sub-problems of $O(L)$ to $N_{sel}$, i.e. if $j$=$i$-$q$ yields the minimum value for $O(L)$ in Eq. \ref{opt_i_eq}, add $j$ to $N_{sel}$. Repeat these steps for $O(j)$, and trace all the way back to reach the boundary sub-problems. At the end, we will have $N_{sel} = \{m_1,m_2,...,m_{K-1}\}$. Having $N_{sel}$, the $i^{th}$ sub-set of nodes, $V_i$, corresponding to the $i^{th}$ part in DLGP problem is obtained using the following equation:
\begin{equation}
\label{gen_Vi_eq}
{\small
V_i = 
\begin{cases}
 \{v \in V | ~0 < \Lambda(v) \leq m_1 ~ \}   ~~~~~~~:   & i=1\\
\{v \in V | ~m_{i-1} < \Lambda(v) \leq m_i ~ \}   ~~:  & ~ 1 < i <K ~\\
 \{v \in V | ~\Lambda(v) > m_{K-1} ~ \}     ~~~~~~~~:   & i=K 
\end{cases}
}
\end{equation}
in which, $\Lambda(v)$ returns the level of node $v$. Algorithm 1 shows the pseudo code of DLGP. Complexity of line 1 and also line 4 are $O(m+n)$, where $n$ is the node count and $m$ is the edge count. Complexity of lines 7-8 is $O(p \times L)$ based on Eq. \ref{opt_i_eq}. Complexity of lines 10-11 is $O(n)$, because we go through all nodes only one time and put them in a part based on Eq. \ref{gen_Vi_eq}. Therefore, the overall complexity of the DLGP algorithm is $O(m+n)$.

\begin{figure}[t]
        \centering
         \begin{subfigure}[!t]{0.275\textwidth}
                \centering
                \includegraphics[width=\textwidth]{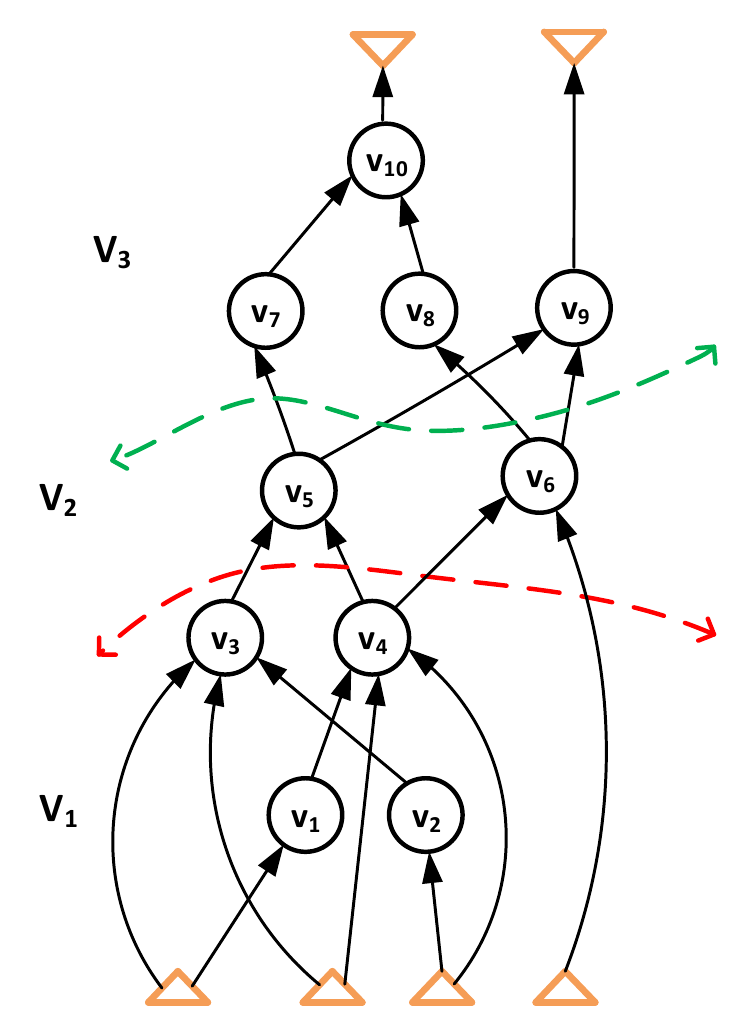}
                \caption{}
                \label{Example2_a_fig}
        \end{subfigure}
        \centering
         \begin{subfigure}[!t]{0.15\textwidth}
                \centering
                \includegraphics[width=\textwidth]{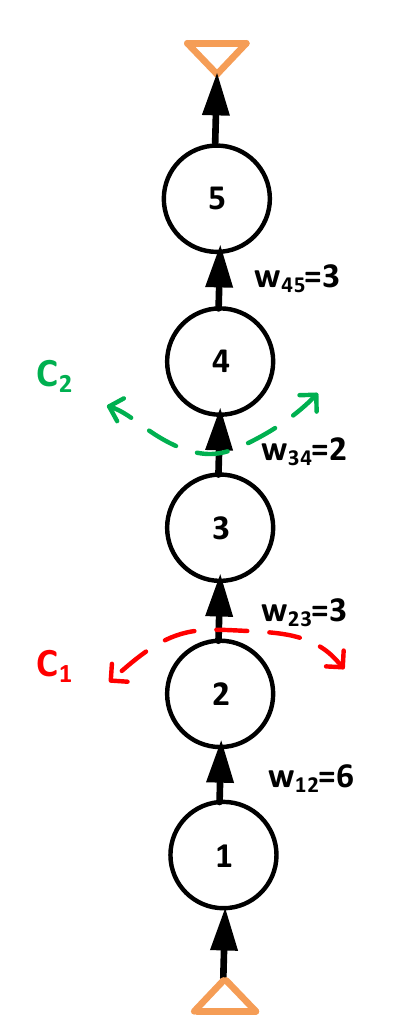}
                \caption{}
                \label{Example2_b_fig}
        \end{subfigure}
        \caption{(a) Graph of Example 2, (b) Corresponding weighted directed chain graph. Cuts $C_1$ and $C_2$ generate three parts $V_1$, $V_2$, and $V_3$ with minimizing the inter-part net weight.}
        \label{Example2}
\end{figure}

\textit{Example 2}: For the graph shown in Fig. \ref{Example2_a_fig} with depth $L$=$5$, by having $p$=$2$, $K$=$3$, and using \textit{hyper-edges} for calculating weights, the corresponding weighted directed chain graph will be as shown in Fig. \ref{Example2_b_fig}. Using the DLGP algorithm, the selected levels will be $N_{sel} = \{ 2,3\}$, and the sub-set of nodes corresponding to optimal parts will be $V_1=\{ v_1,v_2,v_3,v_4\}$, $V_2 = \{ v_5,v_6\}$, and $V_3 = \{ v_7, v_8, v_9, v_{10} \}$. Please note that since \textit{hyper-edges} are used, the edge weights for the weighted directed chain graph will be $\{6,3,2,3\}$ as shown in Fig. \ref{Example2_b_fig} instead of $\{7,4,4,3\}$, which is for the case of using regular edges in weight calculations.
\begin{algorithm} [t]
\caption{DLGP}
\DontPrintSemicolon 
\KwIn{$G=(V,E)$, \\
$p$: \text{constraint on depth}, \\
a mapping function $\Lambda$ which returns level of a node in $G$.\\
}
\KwOut{An optimal set  $P = \{V_1,V_2,..., V_K\}$ of parts.}
$L$ = $Compute\_Graph\_NodeDepth(G)$. 

 \If {$p \geq L$}{
 	\textbf{return} $P=\{V\}$
 }
 Generate the weighted directed chain graph $C=(U,F,w)$.
 
 \For {$i$=1; $i \leq p$; i++}{
 	$OPT(i) = 0$
 }
 
 \For {$i$=1; $i \leq L$; i++}{
 	Find $O(i)$ and calculate its value, $OPT(i)$, using Eq. \ref{opt_i_eq}.
 }
 
 Find $N_{sel} = \{m_1,m_2,...,m_{K-1}\}$ by tracing back from the $O(L)$ solution.
 
 \For {$i$=1; $i \leq K$; i++}{
 	Find $V_i$ using Eq. \ref{gen_Vi_eq}
 }
 
\Return{$P=\{ V_1,V_2,...V_K \}$}.\;
\label{DLGP_alg}
\end{algorithm}
\section{Reducing Path Balancing Overhead in SFQ Circuits}
\label{sec:balancing_sec}
Evaluation of SFQ gates is destructive with respect to any internal state (loop current state) of the gate and any incoming input pulses. In other words, after an SFQ gate receives a clock pulse to produce its output, any stored internal state of the gate is destroyed and the input pulse is consumed. For example, if a 2-input AND gate receives a 1 pulse on its $in_1$ before the clock signal arrives, it will store the said input pulse as a persistent current  in one of its internal loops. Next, if the gate receives a second 1 pulse on its $in_2$ again before the clock comes, it will store this input value in a second internal loop as a persistent current. Finally, when the clock input to the gate arrives, both loop currents will reset (revert back to the other direction of current flow) and an output pulse is reproduced to signify the 1 output for the AND gate. 
Now consider a situation in which $in_1$ arrives in clock cycle 1 whereas $in_2$ arrives in clock cycle 2. One would expect that the AND gate will produce a 0 at the end of clock cycle 1 and a 1 at the end of clock cycle 2. However, this is not the case. In SFQ logic, the AND gate will receive and consume input pulse on $in_1$ during clock cycle 1 producing a 0 output and then it will receive and consume input pulse on $in_2$ in cycle 2, again producing a 0 output. This is precisely why the full path balancing method is employed to make sure that the AND gate will produce the correct 1 pulse output at the end of clock cycle 2. 

Our key observation is that to avoid path balancing DFFs, all we have to do is to make sure that the producer of the 1 pulse on $in_1$ produces that same pulse in both clock cycles 1 and 2. Therefore the AND gate will produce a wrong value of 0 at the end of clock cycle 1 but the correct value of 1 at the end of clock cycle 2. So, as long as we initiated new data toward the AND gate with a slow clock frequency which is half of the fast clock frequency used to clock the AND gate, then the AND gate will produce the correct output at multiples of the slow clock. 

To the best of our knowledge, this is the very first paper that makes this observation and uses it to effectively eliminate path balancing DFFs inside an SFQ logic circuit although the method comes at the expense of using two different clocks (micro and macro clocks) and a number of NDRO (output-replicating or repeating) DFFs. These NDRO DFFs are read by micro clock and are written by macro clock. Since they are being read by micro clock, in each cycle of the micro clock the correct pulses will be re-generated and put on the primary inputs of each part. Please note that DRO DFFs cannot be used here, because the read operation in DRO DFFs is destructive, hence, they cannot re-generate the correct pulses for $p$ times.

As explained above and in Section \ref{PathBalancingSubSubSec}, for correct operation of SFQ circuits, full path balancing is required. One way of addressing the path balancing problem is to add as many DFFs as required to remove any differences among levels of inputs to any SFQ gate. This approach is called \textit{Full Path Balancing (FPB)}. In \cite{katam2017desig_isec}, it is suggested to apply the standard retiming algorithm \cite{leiserson1991retiming} after a heuristic FPB algorithm to minimize the number of path-balancing DFFs (called FPB+retiming). 
In spite of this algorithm, our experiences show that the FPB+retiming will add a large number of DFFs to the circuit which can dominate the original gate count in the network even considering the fact that the area cost of an SFQ DFF is somewhat less than the area cost of say 2-input SFQ AND gate  \cite{RSFQLib_CF}. In this section, we will show how DLGP algorithm helps solving this problem.

\begin{figure}[t]
        \centering
         \begin{subfigure}[!t]{0.32\textwidth}
                \centering
                \includegraphics[width=\textwidth]{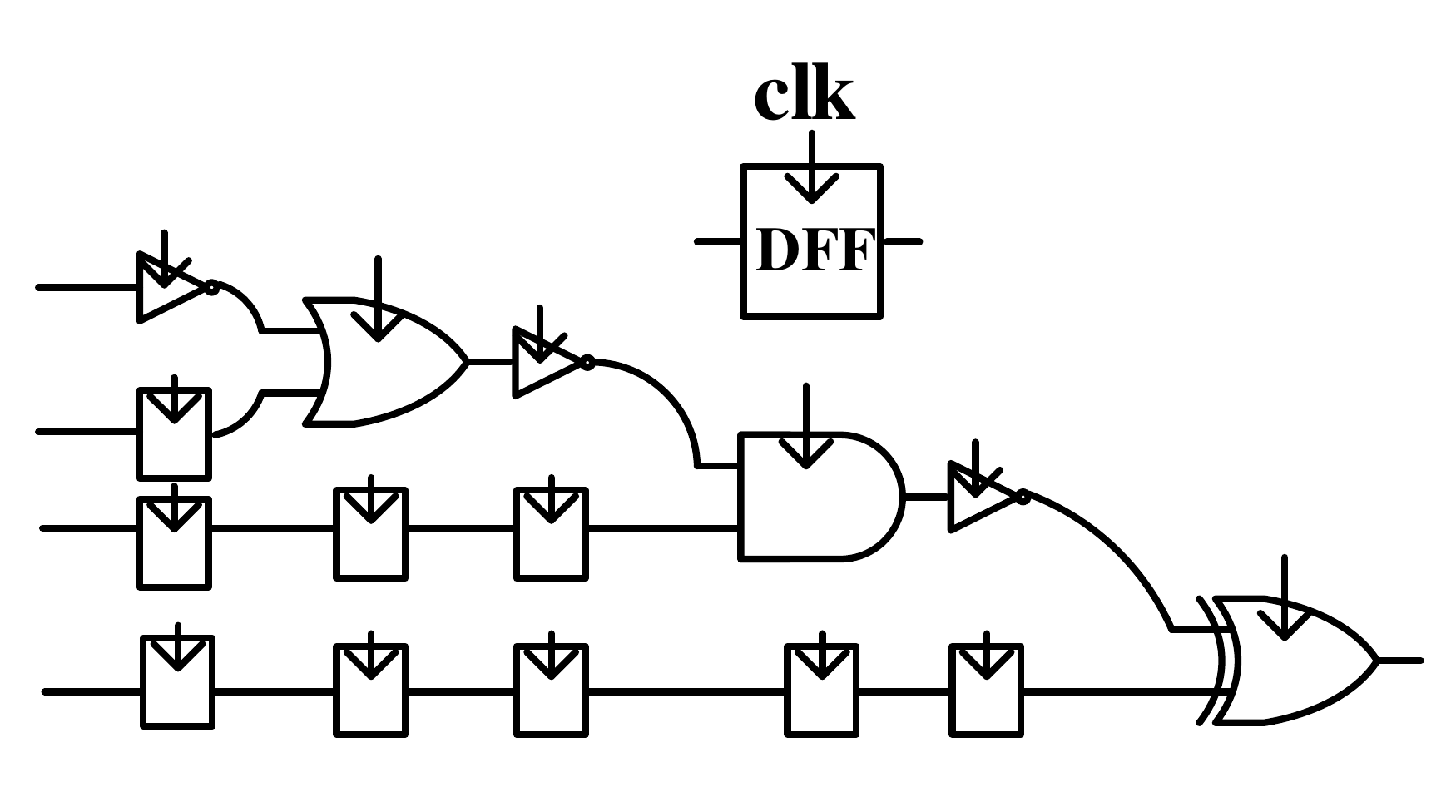}
                \caption{}
                \label{fully_balanced_fig}
        \end{subfigure}
         \centering
         \begin{subfigure}[!t]{0.32\textwidth}
                \centering
                \includegraphics[width=\textwidth]{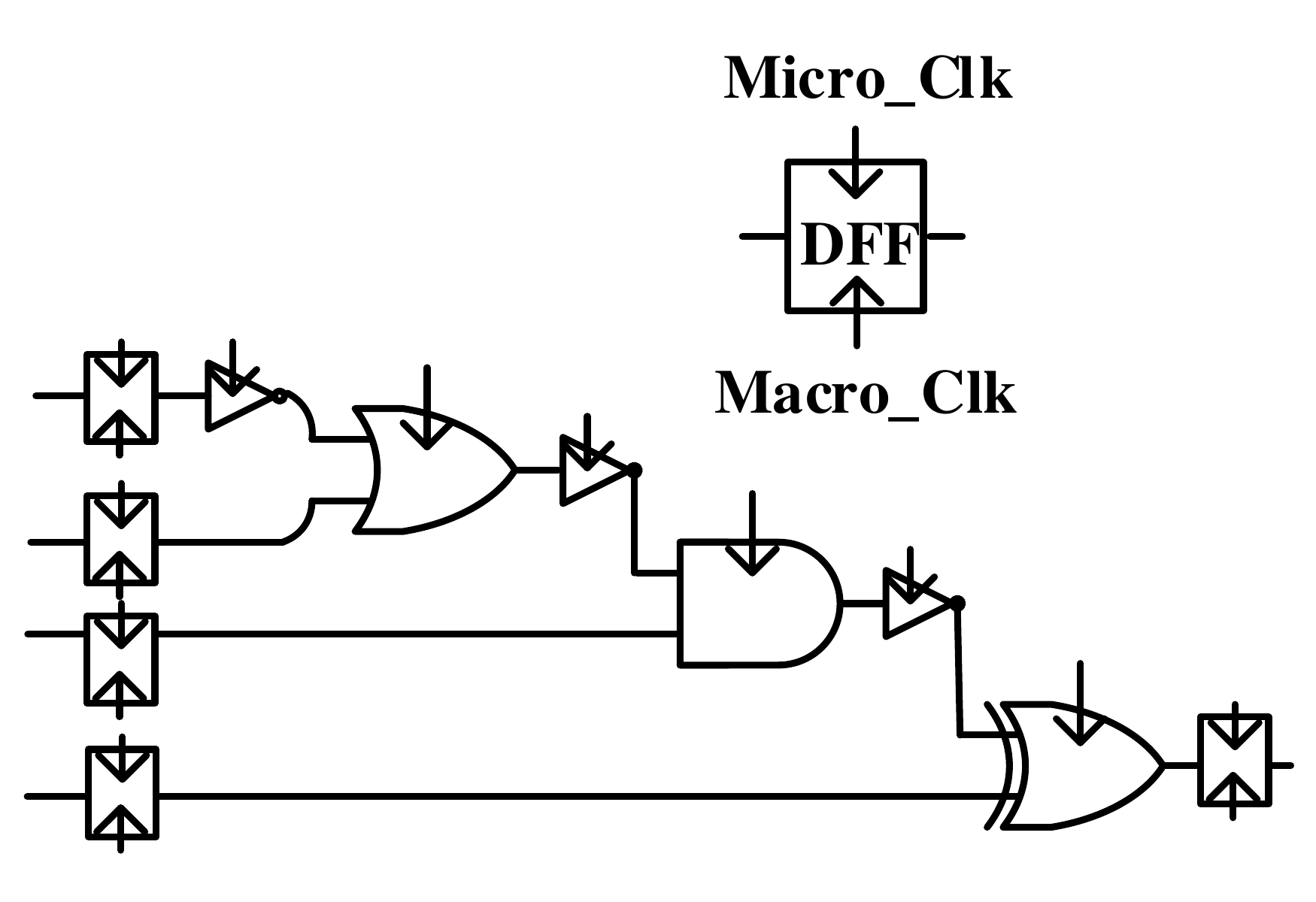}
                \caption{}
                \label{DLGP_balanced_fig}
        \end{subfigure}
        \caption{An example of (a) full path balancing (FPB), and (b) DLGP-based dual clocking method (DCM) (with $p$=$6$) for SFQ circuits. FPB uses DRO DFFs whereas the DCM uses NDRO DFFs. FPB requires 9 DRO DFFs and DCM requires 5 NDRO DFFs.}
        \label{Fully_vs_DLGP}
\end{figure}

As hinted earlier, we propose to use a fast \textit{micro} and a slow \textit{macro} clock and to use the DLGP algorithm to minimize the aforementioned overheads of FPB. Thanks to the DLGP algorithm, it is possible to divide the corresponding graph of a given SFQ circuit into a few depth-limited parts and add NDRO DFFs only on the hyper-edges which are cut by various part boundaries. These DFFs will pass values that go from one part to the other one with the macro clock, while  gates inside each part operate with the micro clock. Since the DLGP algorithm guarantees giving the minimum total cut weight, the number of inserted NDRO DFFs in the circuit will be minimized. Furthermore, since the NDRO DFFs which are placed at the inputs of each part are also clocked by the micro clock to continuously reproduce their outputs, there is no need to add any path balancing DFFs inside each part. The resulting SFQ circuit is thus functionally pipelined allowing a number of $K$ data instances to exist in the circuit at the same time, each being operated in the corresponding part of the $K$-part circuit. The number of parts will thus affect the total operational throughput of the circuit (when there are no pipeline installs). Fig. \ref{Fully_vs_DLGP} shows FPB, and DLGP-based Dual Clocking Method (DCM) for an example circuit. As seen, the DLGP-based DCM requires 4 fewer number of DFFs compared with FPB. Note that although NDRO DFFs are more expensive than the DRO DFFs, the reduction in total DFF count far outweights this difference in element cost. See experimental results. 

In our DLGP-based DCM, the depth of each part is at most $p$. 
Since  gates in each part are evaluated by the micro clock and the speed of micro clock is $p$ times faster than the macro clock, between two consecutive edges of the macro clock, $p$ different values will hit the inputs of DFFs at the inter-part boundaries. If every such value reaches the NDRO DFFs, this will cause wrong values to be stored in the NDRO DFFs, which will be passed to the next part. Indeed we want that only the value which is generated in the last cycle of the micro clock is written into the NDRO DFFs, because only this value is valid. This issue can easily be addressed using the following pulse-repeating gate and by ensuring that the macro clock is synchronized with the micro-clock but has a clock frequency which is $p$ times slower. 
\begin{figure}[t]
\centering
\includegraphics[width=0.23\textwidth]{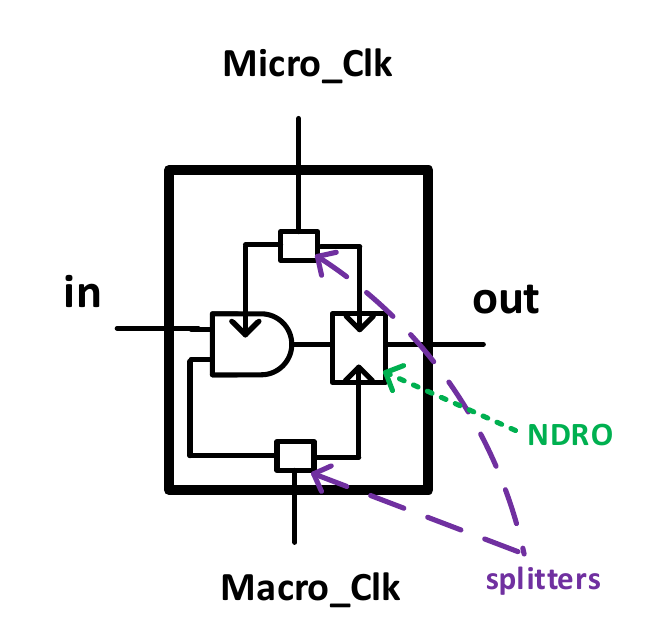}
\caption{Pulse-repeating gate: an SFQ gate consists of an NDRO DFF, an AND gate, and two splitters to give the macro and micro clocks to these gates.}
\label{P_C}
\end{figure}
\subsubsection{Pulse-Repeating Gate} \label{Pulse_Cloning_Gate}
We have invented an SFQ \textit{pulse-repeating gate} shown in Fig. \ref{P_C} to address the aforesaid challenges in the DLGP-based DCM. In this gate, an AND gate is added to the input of the NDRO DFF gate. One of the inputs of this AND gate is connected to the macro clock which is ``0" while the correct value is not generated on  the ``in" port. Therefore, it will pass a ``0" to the input of the NDRO DFF as desired. Only in the last cycle of the micro clock the AND gate will be transparent and pass the valid value to the NDRO DFF. In addition, due to usage of NDRO DFF, inputs of each part will be re-generated (repeated) at each cycle of the micro clock.  

Before passing these circuits to the DLGP-based DCM, they are passed to the technology mapping engine of ABC \cite{synthesis2011abc} and its default optimization are performed on them. After that, mapped circuits are passed to the DLGP algorithm to find the optimum places for inserting pulse-repeating gates. After this step, splitter insertion and balancing of Primary Outputs (POs) are performed.
Algorithm 2 shows the pseudo code for DLGP-based DCM. In line 1, the given circuit is mapped using the technology mapping engine of ABC. In line 2, the optimal parts are determined. In line 3, the pulse-repeating gates are inserted on the hyper-edges which are cut. Line 4 takes care of balancing POs, and finally, line 5 inserts the required splitters. 
\section{Experimental Results}
\label{exper:sec}
\begin{algorithm} [t]
\caption{DLGP-based Dual Clocking Method}
\DontPrintSemicolon 
\KwIn{a graph $G=(V,E)$ corresponding to the input circuit, \\
$p$: \text{constraint on depth}}
\KwOut{A timing-correct circuit represented by graph $G'=(V',E')$}
$G_m$ = $Technology\_Mapping(G)$
 
$Parts=\{V_1,V_2,...,V_k\}$ = $DLGP(G_m,p)$

$Pulse\_Repeating(G_m,Parts)$

$PO\_Balancing(G_m)$ 

$G' = Splitter\_Insertion(G_m)$

\Return{$G'$}\;
\label{algo:DLGP_PB}
\end{algorithm}
We implemented  the DLGP algorithm inside ABC. An SFQ library of gates as in \cite{RSFQLib_CF} is used. This library consists of the following gates: \textit{and2} with 12 JJs, \textit{or2} with 8 JJs, \textit{xor2} with 8 JJs, \textit{DFF} with 7 JJs, \textit{splitter} with 3 JJs, \textit{JTL} with 2 JJs, and \textit{not} with 9 JJs. A few benchmark circuits from ISCAS \cite{hansen1999unveiling}, EPFL \cite{EPFL_bench}, and some other arithmetic circuits are chosen to test the effectiveness of the proposed algorithm in reducing the overhead of FPB. 
\begin{table*}[th]
\scriptsize
  \centering
  \caption{Experimental results for DLGP-based DCM, Baseline1 (FPB), and Baseline2 (FPB + retiming). Area is in $mm^2$ and run-time is in \textit{sec}. For DLGP-based dual clocking method (DCM), two cases of $p$=$10$, and $p$=$5$ are considered.}
    \begin{tabular}{ccccccccccccc}
    \toprule
     & \multicolumn{4}{c}{Area} & \multicolumn{4}{c}{\#JJs}& \multicolumn{4}{c}{Run-Time} \\
   circuits  &  Baseline1 & Baseline2 & DLGP(5) & DLGP(10)  &  Baseline1 & Baseline2 & DLGP(5) & DLGP(10) & Baseline1 & Baseline2 & DLGP(5) & DLGP(10) \\
    \toprule

i10	  &120.19	 &88.15	 &80.23	 &53.57	 &170126	&125626	  &124138	&81713	 &3.66	 &7.24	 &0.248	 &0.227  \\
\midrule
c1908	&10.57	 &8.26	 &8.65	 &5.9	  &14783	&11575	  &13169	&8739	 &0.038	 &0.109	  &0.013  &0.013  \\
\midrule
c1355	&6.9	 &5.66	 &5.95	 &4.35	  &9434	    &7718	   &8739	&6149	 &0.041	 &0.057	  &0.033	&0.033  \\
\midrule
c432	&8.7	&7.3	&6.9	&4.7	 &12288	    &10368	   &10734	&7124	 &0.0248	&0.053	 &0.008	  &0.008  \\
\midrule
c880	&13.11	 &8.98	&9.51	&6.28	 &18561	    &12831	   &14658	&9483	  &0.045	&0.0135	  &0.008	&0.008  \\
\midrule
c3540	&33.2	 &21.48	 &28.07	&17.83	  &47253	&30937	   &43437	&26897	   &0.302	&0.549	  &0.028	&0.028  \\
\midrule
voter	&311.83	  &258.92	&324.68	 &249.69	&447044	 &373559	&496238	 &374718	&28.42	&33.67	&0.785	 &0.755  \\
\midrule
int2float	&6.67	&4.79	&5.06	 &3.43	 &9539	 &6919	 &7770	 &5140	 &0.014	  &0.034	&0.006	 &0.006  \\
\midrule
ADD16	&3.72	&3.47	&3.75	&2.63	&5411	&5061	&5726	&3911	&0.012	&0.0016	&0.0093	 &0.0094  \\
\midrule
MULT16	&169.08	 &39.61	 &107.61	&58.64	&236319	 &56489	&171787	 &92647	 &5.63	&25.63	 &0.134 	&0.134  \\
	\bottomrule
    \end{tabular}%
  \label{exp_table}%
\end{table*}%
\begin{table}[th]
\scriptsize
  \centering
  \caption{Comparing DFF count for DLGP-based DCM and two baselines. }
    \begin{tabular}{ccccc}
    \toprule
     & \multicolumn{4}{c}{\#DFFs} \\
   circuits  &  Baseline1 & Baseline2 & DLGP(5) & DLGP(10)   \\
    \toprule

i10	  &12832	&8382	&3219	&1872 \\
\midrule
c1908	&1004	&683	&282	&144  \\
\midrule
c1355	&614	&442	&193	&119  \\
\midrule
c432	&847	&655	&224	&118  \\
\midrule
c880	&1345	&772	&362	&187 \\
\midrule
c3540	&2848	&1220	&776	&282  \\
\midrule
voter	&18491	&11114	&7204	&3732  \\
\midrule
int2float	&539	&277	&117	&39  \\
\midrule
ADD16	&235	&200	&104	&50  \\
\midrule
MULT16	&21373	&3390	&4460	&2111 \\

	\bottomrule
    \end{tabular}%
  \label{exp2_table}%
\end{table}%

Tables \ref{exp_table}, \ref{exp2_table} show experimental results for DLGP-based DCM with two values for $p$, 10 and 5. For a better comparison, two baselines are considered; \textit{Baseline1} is the FPB, and \textit{Baseline2} is the FPB+retiming algorithm. As seen, DLGP-based DCM provides substantial savings in total number of Josephson junctions (\#JJs), area,  run-time, and DFF count (\#DFFs). \textit{Area} and \textit{\#JJs} are area and JJ count for gates, DFFs and splitters. The overhead of AND gates (in pulse-repeating gates) and second clock in DLGP-based DCM are considered in the experimental results of DLGP. \#DFFs for DLGP includes NDRO DFFs inserted to the boundary of parts and DRO DFFs used for PO balancing. In DLGP-based DCM and both baselines, the cut-based technology mapping of ABC (command ``map") which involves a delay optimization pass followed by a few area optimization passes is employed. Other than what mentioned so far, no other optimization function is used.

DLGP-based DCM for $i10$ MCNC benchmark circuit consumes $1.54 \times$ and $1.37 \times$ fewer \#JJs compared with Baseline1 and Baseline2, respectively when $p$=$10$. For the same, DLGP-based DCM provides $2.24 \times$ and $1.65 \times$ improvements on total area, and  $6.85 \times$ and $4.47 \times$ improvements on \#DFFs compared with Baseline1 and Baseline2, respectively. For $p$=$5$ the amount of improvements are less than these values. For example, for the same circuit, the saving of $2.24 \times$ and $1.65 \times$ on area is reduced to $1.50 \times$ and $1.09 \times$, and the saving on \#DFFs is decreased from $6.85 \times$ and $4.47 \times$ to $3.99 \times$ and $2.60 \times$ all compared with Baseline1 and Baseline2, respectively. 

On average for all 10 benchmark circuits, the saving on area, \#JJs, and \#DFFs for DLGP-based DCM when $p$=$10$ is 89\%, 77\%, and $7.7 \times$, respectively over Baseline1 and 30\%, 23\%, and $4.26 \times$, respectively over Baseline2. The reason behind not seeing the huge saving of \#DFFs in the total area is the overhead of second clock and also the AND switches used in the pulse-repeating gates. DLGP-based DCM also decreases the run-time significantly. For example for $c432$ benchmark circuit, the run-time is decreased by $3.10 \times$ and $6.63 \times$ when $p$=$5$ compared with Baseline1 and Baseline2, respectively. The main reason behind larger run-time for baselines is requirement of inserting many DFFs plus performing retiming, which both are slow processes specially for large benchmark circuits. We tried to extract experimental results for larger benchmark circuits (larger than \textit{voter} with 13758 nodes, 1002 IOs, 27516 edges, and 13758 cubes which is already reported in Tables \ref{exp_table}, \ref{exp2_table}) such as \textit{log2} and \textit{hypotenuse} \cite{EPFL_bench}, but the memory of our system (64GB RAM) was not enough for finishing DFF insertion and retiming steps of Baseline1 and Baseline2, and the processes were killed after 30+ minutes. 
\begin{figure}[b]
\centering
\includegraphics[width=0.43\textwidth]{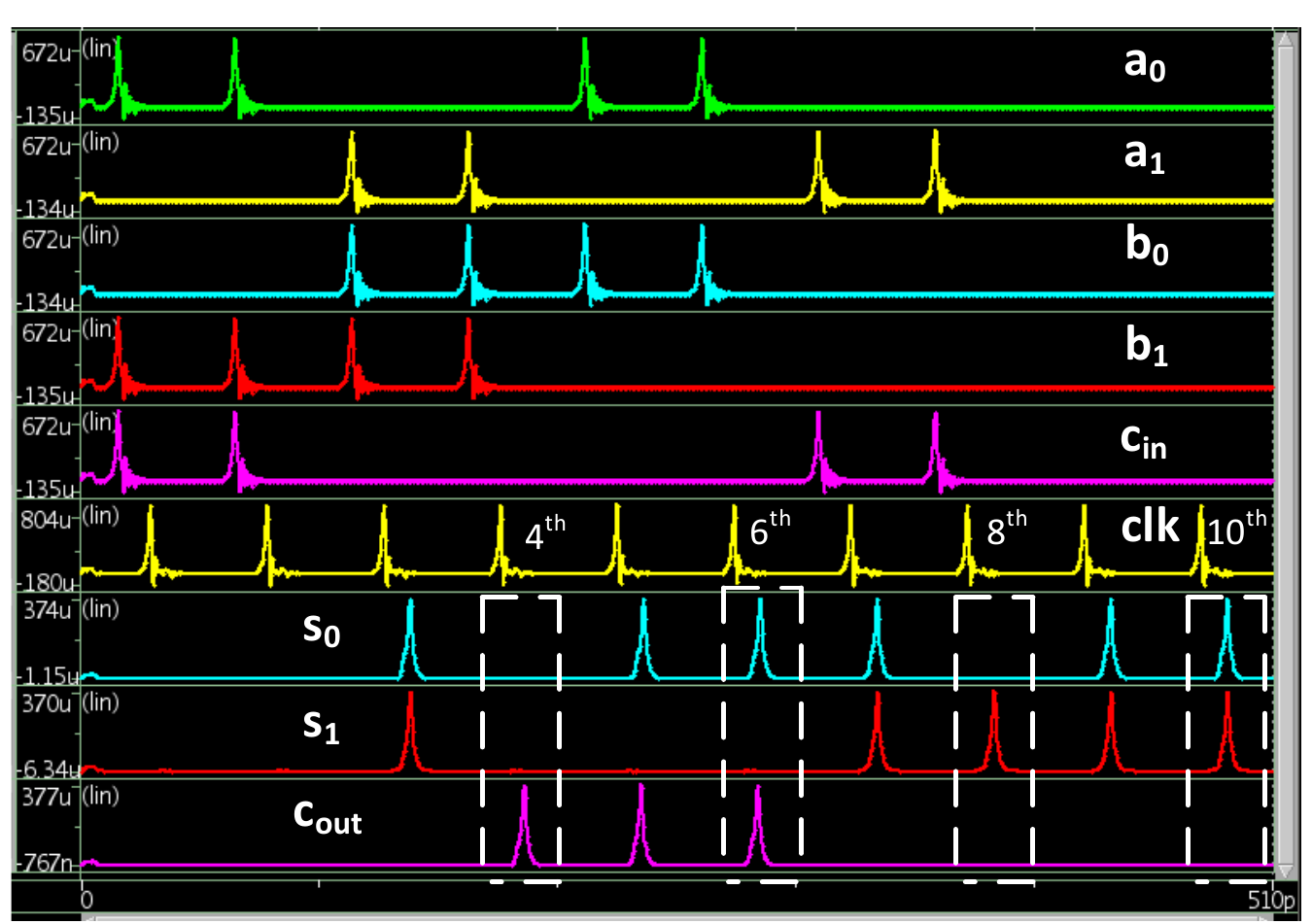}
\caption{Simulation results for a 2-bit Kogge-Stone adder (KSA2) generated by DLGP-based DCM given $p$=$2$. \textit{clk} refers to the fast clock. Four sets of random inputs are applied: $a_0$=1010, $a_1$=0101, $b_0$=0110, $b_1$=1100, $c_{in}$=1001. The correct outputs are: $S_0$=0101, $S_1$=0011, $C_{out}$=1100, which are generated every 2 clock cycle (fast clock).}
\label{KSA2_sim_fig}
\end{figure}

To verify the correct operation of the circuits, we simulated a 2-bit Kogge-Stone Adder (KSA2) generated by DLGP-based DCM given $p$=$2$.  Four sets of random values as shown in Fig. \ref{KSA2_sim_fig} are applied to the inputs and for all of them the correct outputs are generated. Please note that since $p$=$2$, the inputs are repeated and there are 2 copies of each input. 

These savings come in one expense; the peak throughput of circuits generated by DLGP-based DCM will be roughly $p \times$ less than FPB method. This is because the effective frequency of these circuits is the same as macro clock compared with the frequency of circuits generated by FPB which is micro clock. Note that the actual throughput is typically much less than the peak throughput (due to instruction data dependencies, program branches, etc.); so some throughput loss may be acceptable. In addition, due to the following property of SFQ circuits, the throughput loss will be less than $p \times$ after place-and-route; in SFQ circuits, the delay of interconnects are typically larger than the delay of gates, hence, the longest interconnect usually determines the worst case delay. Therefore, since DLGP-based DCM reduces the total gate count significantly, it will help reducing the length of the longest interconnect, resulting in having faster local clock frequency. This helps gaining some of the lost throughput in DLGP-based DCM. For example, the throughput loss for ISCAS c432 circuit generated by DLGP-based DCM ($p$=$5$) after place-and-route is reduced to $4.3 \times$ compared with FPB. A more advanced wire-routing method such as what presented in \cite{kito2018fast} can help reducing this gap further.
\section{Conclusion} \label{sec:conclusion}
This paper introduces a new graph partitioning problem called Depth-bounded Levelized Graph Partitioning (DLGP). In DLGP, there is a depth constraint on the resulting sub-graphs of each part. We showed that by transforming the DLGP problem into a Depth-bounded Chain Graph Partitioning (DCGP) problem, an optimal solution which minimizes the total cut set is achieved using the Dynamic Programming algorithm. It is shown that DLGP algorithm can be applied to SFQ circuits for reducing the path balancing overheads. Experimental results show that if the depth constraint is equal to 5, this overhead reduction is as high as $1.38 \times$ in terms of JJ count, and $4.79 \times$ in terms of DFF count.

\ifCLASSOPTIONcaptionsoff
  \newpage
\fi
\bibliographystyle{IEEEtran}
\bibliography{IEEEabrv,template}

\end{document}